\documentclass[letterpaper,english,aps, prx, superscriptaddress, amsfonts, amssymb, amsmath, a4paper, reprint]{revtex4-2}
\usepackage[T1]{fontenc}
\usepackage[latin9]{inputenc}
\usepackage{color}
\usepackage{nicefrac, xfrac}
\usepackage{babel}
\usepackage{array}
\usepackage{booktabs}
\usepackage{textcomp}
\usepackage{multirow}
\usepackage{amsmath}
\usepackage{amssymb}
\usepackage{graphicx}
\usepackage{upgreek}
\usepackage[unicode=true,pdfusetitle, bookmarks=true,bookmarksnumbered=false,bookmarksopen=false,
 breaklinks=false,pdfborder={0 0 1},backref=false,colorlinks=true]
 {hyperref}

\makeatletter

\providecommand{\tabularnewline}{\\}

\makeatother

\begin{document}
\global\long\def\ket#1{\left|#1\right\rangle }%
\global\long\def\bra#1{\left\langle #1\right|}%
\global\long\def\ketbra#1#2{\ket{#1}\bra{#2}}%

\global\long\def\braket#1#2{\langle#1|#2\rangle}%
\global\long\def\bravket#1#2#3{\langle#1|#2|#3\rangle}%
\global\long\def\avg#1{\langle#1\rangle}%

\global\long\def\abs#1{|#1|}%

\global\long\def\pa{\partial}%

\global\long\def\hhup{\Uparrow}%
\global\long\def\hhdown{\Downarrow}%
\global\long\def\hhud{\hhup\!\hhdown}%
\global\long\def\hhdu{\hhdown\!\hhup}%
\global\long\def\lhup{\uparrowtail}%
\global\long\def\lhdown{\downarrowtail}%

\global\long\def\elup{\uparrow}%
\global\long\def\eldown{\downarrow}%
\global\long\def\elud{\updownarrows}%
\global\long\def\eldu{\downuparrows}%

\title{Magneto-optics of a charge-tunable quantum dot - Observation of a
negative diamagnetic shift}
\author{G. Peniakov}
\email[Corresponding author: ]{giora.peniakov@uni-wuerzburg.de}
%\thanks{These authors contributed equally to this work.}
\affiliation{The Physics Department and the Solid State Institute, Technion\textendash Israel
Institute of Technology, 3200003 Haifa, Israel}
\affiliation{Julius-Maximilians-Universit{\"a}t W{\"u}rzburg, Physikalisches Institut, Lehrstuhl f{\"u}r Technische Physik, Am Hubland, 97074 W{\"u}rzburg, Germany}
\author{A. Beck}
\thanks{These authors contributed equally to this work.}
\affiliation{The Physics Department and the Solid State Institute, Technion\textendash Israel
Institute of Technology, 3200003 Haifa, Israel}
\author{E. Poem}
\thanks{These authors contributed equally to this work.}
\affiliation{Department of Physics of Complex Systems, Weizmann Institute of Science,
7610001 Rehovot, Israel}
\author{Z-E. Su}
\affiliation{The Physics Department and the Solid State Institute, Technion\textendash Israel
Institute of Technology, 3200003 Haifa, Israel}
\author{B. Taitler}
\affiliation{The Physics Department and the Solid State Institute, Technion\textendash Israel
Institute of Technology, 3200003 Haifa, Israel}
\author{S. H{\"o}fling}
\affiliation{Julius-Maximilians-Universit{\"a}t W{\"u}rzburg, Physikalisches Institut, Lehrstuhl f{\"u}r Technische Physik, Am Hubland, 97074 W{\"u}rzburg, Germany}
\author{G.W. Bryant}
%\affiliation{National Institute of Standards and Technology, 100 Bureau Drive, Gaithersburg, Maryland 20899, USA}
\affiliation{Nanoscale Device Characterization Division, National Institute of Standards and Technology, Gaithersburg, Maryland 20899, USA}
%\affiliation{Nanoscale Device Characterization Division, Joint Quantum Institute, National Institute of Standards and Technology, Gaithersburg, Maryland 20899;  University of Maryland, College Park, Maryland 20742, USA}
\affiliation{Joint Quantum Institute, National Institute of Standards and Technology and University of Maryland, College Park, Maryland 20742, USA}
\author{D. Gershoni}
\affiliation{The Physics Department and the Solid State Institute, Technion\textendash Israel
Institute of Technology, 3200003 Haifa, Israel}
\date{\today}
\begin{abstract}
We present magneto-optical studies of a self-assembled semiconductor
quantum dot in neutral and positively charged states. The diamagnetic shifts and Zeeman splitting of many well-identified optical transitions are precisely measured.
Remarkably, a pronounced negative diamagnetic shift is observed for spectral lines resulting from a doubly positively charged excitonic complex.
We use the Hartree - Fock approximation for describing the direct Coulomb and exchange interactions between the quantum dot confined carriers in various configurations. 
A simple harmonic potential model, which we extend to capture the influence of an externally applied magnetic field in Faraday configuration, is then used to 
quantitatively account for all the measured diamagnetic shifts. 
We show that the negative shift is due to the change in the hole-hole exchange interaction energy induced by the magnetic field. 
Using this model  and the measured shifts we extract the dielectric constant of the quantum dot material and get a decent estimate of the quantum dot dimensions. 
Further, the measured Zeeman splitting of the various spectral lines are also explained by a simple model using algebraic sums and differences of the $g$-factors of 
the confined charge carriers in their respective first and second discrete energy levels. 
Finally, the obtained values of the electronic $g$-factor and that of the dielectric constant are independently used to determine the effective composition (x)
of the ternary  In$_{x}$Ga$_{1-x}$As quantum dot. Both agree to within the experimental uncertainties.   
\end{abstract}
\maketitle

\section{Introduction}

Self-assembled Quantum Dots (QDs) in semiconductors form a well-known
platform for quantum technologies. They have proven to be the best
contemporary single-photon sources \citep{Dekel2000,Michler.2000,Tomm2021,Santori2001,cylu2021},
while providing an excellent interface between anchored spin qubits
and \textquotedblleft flying\textquotedblright{} photon qubits \citep{Cogan2022}. Much
progress has been made in controlling confined-spin qubits in QDs
\citep{Awschalom2008,Yamamoto2008,Kodriano2012,Schwartz2015},
entangling them with photons \citep{Akopian2006,Lukin2010,Ohno2012,Gao2012,DeGreve2012,Schaibley2013}\textcolor{green}{}
, and allowing deterministic generation of long strings of entangled photons
\citep{cluster,Peniakov2020,Cogan2023}. \textcolor{green}{} In addition,
QDs still provide a convenient platform for studying many-body complexes composed of multiple confined carriers. Interesting properties
of such complexes include the relative interactions between the constituent
particles, the form of their spatial wavefunctions, and their response
to externally applied fields. In particular, an externally
applied magnetic field removes the Kramers' degeneracy and causes the optical transitions to
energetically shift. The first interaction, known as the Zeeman-interaction, depends linearly on the field strength, while the second interaction known as the \textit{diamagnetic shift}, depends quadratically on the field strength \cite{Ivchenko2005}. 
Modeling these shifts in confined systems and in particular in semiconductor nanostructures is still a subject of many research efforts
\citep{Fu2010,Glazov2007,Shin:15,Walck1998,Schulhauser2002}.

We present here a magneto-optical study, mainly in Faraday configuration, of a semiconductor QD with very well-identified optical transitions between various few-confined-carriers configuration in different charge states. 
In particular, we focus our attention on the magneto optical properties of the doubly, positively-charged excitonic complex $X^{+2}$.

The $X^{+2}$ can be intuitively described as a confined multi-carrier configuration containing three heavy holes and one
electron. After a radiative recombination of an electron-hole pair,
the QD remains with two holes, one in the lowest energy level and one in the second level. These two holes may form either
three spin triplet states or one spin singlet state. Our work was
spurred by noticing an anomaly in the diamagnetic shifts of the optical
transitions into the singlet state, labeled $X_{S_{0}}^{+2}$, which we
found to be {\bf negative}. 
In the effort of understanding this phenomenon,
we found that the $X^{+2}$ excitonic transitions form an excellent
platform for studying the electron-hole and hole-hole exchange interactions and
their dependence on an externally applied magnetic field. 
Using the Hartree-Fock approximation and 
a simple two dimensional (2D), cylindrically symmetric,  harmonic-oscillator model for describing the QD spatial potential, we quantitatively describe all the measured diamagnetic shifts and in turn determine the QD's average dielectric constant and its dimensions.   
In addition, from the measured Zeeman splittings of the various spectral lines we determine the electron
and hole $g$-factors in their respective two lowest discrete energy levels. We show that the heavy hole $g$-factor switches sign between the first and second confined levels. 

Finally, we show that both the obtained dielectric constant and the electron $g$-factor provide a way to estimate the  In$_{x}$Ga$_{1-x}$As QD's average composition (x). 

The paper is organized as follows. In section II, we briefly describe the experimental system. In section III we present 
full polarization-sensitive magneto-PL measurements displaying the diamagnetic shifts and Zeeman splitting of many
well-identified optical transitions and their polarization selection rules.
We provide a simple model to explain the measured Zeeman splittings of the spectral lines and present in particular the anomalous negative diamagnetic shift of the doubly
positively charged exciton, the $X_{S_{0}}^{+2}$. 
In section IV we discuss in detail the theoretical model that we use to analyze the measured magneto-optics data and in section V 
we compare the model predictions and the measured results. 

\section{Experimental system}

We studied a single In$_{x}$Ga$_{1-x}$As self-assembled QD embedded
in a planar microcavity grown along the $[001]$ direction. The actual dimensions of the studied QD, about 35 nm in diameter and 3-4 nm in height, were estimated from a set of detailed optical studies and comparisons with many-carrier model simulations~\cite{Benny2012}. These structural characterizations were also backed by atomistic simulations~\cite{Zielinski2015}.  
We used an Attocube\textsuperscript{TM}\cite{NISTdisclaimer} closed-cycle cryostat to cool the sample down
to 4 K. A built-in vector magnet enabled the application of a magnetic
field in any desired direction. The emitted photoluminescence (PL)
was collected by a $\times60$ objective. Its polarization was analyzed
by pairs of liquid crystal variable retarders and polarizing beam
splitters, enabling PL polarization projection on any direction in
the Poincar{\'e} sphere. The PL was then spectrally analyzed using an
80 cm double monochromator, providing spectral resolution of $\sim20$~\text{$\mu$}eV.

The QD was optically excited using an above band-gap continuous wave red HeNe laser or
a blue diode laser, emitting at 633 nm or 445 nm, respectively. The excitation
color affects the average charge state of the QD. While red HeNe illumination
results in positive charging, blue diode laser excitation leads to neutral or negative charging
\citep{Benny2012}.

We define the symmetry axis of the QD and the optical beam path 
as the $z$ direction. The $x$ and $y$ axes are
defined along the polarization eigenstates of the QD's bright exciton (BE),
$X_{BE}^{0}$, as explained below.

\section{Results}

\subsection{The neutral bright and dark excitons.} 

The $X_{BE}^{0}$ is an electron-hole pair which can
be expressed in the spin basis $\left\{ \ket{+z}=\ket{\hhup\eldown},\ket{-z}=\ket{\hhdown\elup}\right\} $
with $\hhup$\textbackslash$\hhdown$ and $\elup$\textbackslash$\eldown$
denoting the spin projections of the heavy-hole and electron onto
the z-axis. Since a heavy-hole and an electron have total angular
momenta of $3/2$ and $1/2$, respectively, the angular momentum projection
of a $\ket{\hhup\eldown}$ ($\ket{\hhdown\elup}$) pair along this
axis is +1 (-1) \citep{Poem2007}. Consequently, optical recombination
of the $\ket{\hhup\eldown}$ and $\ket{\hhdown\elup}$ pairs results
in a right-handed ($R$) and left-handed ($L$) circularly polarized
photon emission, respectively. 

The anisotropic electron-hole exchange
interaction in this QD lifts the degeneracy of the above basis by
$\delta_{1}^{1e1h}\thickapprox30~\mu eV$ \citep{Benny2011} thus
forming new eigenstates $\sqrt{2}\ket{\psi_{BE}}_{s,as}=\sqrt{2}\ket{\pm x}=\ket{\hhup\eldown}\pm\ket{\hhdown\elup}$, where the subscript $s$ ($as$) stands for symmetric (anti-symmetric) spin wavefunctions.
Recombination of those excitonic eigenstates results in either horizontal,
$\sqrt{2}H=R+L$, or vertical $\sqrt{2}V=i(R-L)$ rectilinearly polarized  photon
emission, enabling a one-to-one correspondence between the $X_{BE}^{0}$'s
two-level system and the photon's polarization \cite{Benny2011}. We use the cross-rectilinearly polarized components of the $X_{BE}^{0}$ spectral lines for defining 
the $x$ and $y$ axes of our experimental system.  

The dark exciton (DE), $X_{DE}^{0}$, is another electron-hole pair state, but
with parallel spins $\sqrt{2}\ket{\psi_{DE}}_{s,as}=\ket{\hhup\elup}\pm\ket{\hhdown\eldown}.$  In general, the $X_{DE}^{0}$ is optically \textbf{in}active \citep{Schwartz2015},
however, small optical activity of the $X_{DE}^{0}$ was measured
\citep{Schwartz2015a,Don2016,EmmaTutorial2014}\textcolor{green}{}.
Zielinski et al. attributed this activity to a small mixing of the $X_{DE}^{0}$
and $X_{BE}^{0}$ eigenstates  induced by symmetry reduction of the QD potential \citep{Don2016}. 
This naturally occurred mixing is rather small \cite {Schwartz2015} but can be enhanced by externally applying in-plane magnetic field perpendicular to the optical
$z$ axis (Voigt configuration). For a magnetic field direction parallel to the $z$ 
axis (Faraday configuration) no additional mixing is expected, thus
the $X_{DE}^{0}$ remains optically inactive \citep{Bayer2002}\textcolor{green}{}.

%\subsection{Single-carrier $g$-factors\label{subsec:Measuring-single-carrier g-factor}}

The Zeeman interaction between an externally applied magnetic field
and QD confined carriers' spin removes the Kramers' degeneracy between
the confined carriers spin state. The eigenstates  are usually the parallel and anti-parallel
spin directions relative to the direction of the magnetic field. 
The Zeeman interaction depends linearly on the magnetic field magnitude. This dependence is most generally expressed in terms of a $3\times3$ $g$-factor tensor\textcolor{black}{{} \citep{Cardona2010}}.
For simplicity, we assume here
that this tensor is diagonal and have only two different components:
along the symmetry axis ($g_{e}^{z}$ and $g_{h}^{z}$) and perpendicular
to it ($g_{e}^{\perp}$ and $g_{h}^{\perp}$) \textcolor{black}{\citep{Witek2011}}.
Thus in Faraday configuration the Zeeman splitting is given by Eq.~\ref{eq:Zeeman_Hamiltonian}
\begin{equation}
\mathcal{H}=-\mu_{B}g_{e}^{z}B_{z}S_{z}+\frac{1}{3}\mu_{B}g_{h}^{z}B_{z}J_{z}\,\,.\label{eq:Zeeman_Hamiltonian}
\end{equation}
Here, $\mu_{B}$ is the Bohr magneton, $S_{z}$ and $J_{z}$ are the
angular momentum z-projections $\pm\frac{1}{2}$ and $\pm\frac{3}{2}$, of the confined electron and heavy hole respectively
and $B_{z}$ is the magnitude of the magnetic field. 

In the first part of the experiment, we measured the confined electron
and hole $g$-factors tensor components along the z-axis. This was
done by measuring the Zeeman splitting of the $X_{BE}^{0}$ and $X_{DE}^{0}$
under B-field in the $\hat{z}$ direction. Assuming that the absolute
magnitude of the $g$-factors of those transitions are given by the
sum and difference of the absolute magnitudes of the single-carrier
$g$-factors \textcolor{green}{}
\begin{equation}
|g_{BE(DE)}^{z}|=|g_{e}^{z}|\pm|g_{h}^{z}|\label{eq: g_BE)DE)}
\end{equation}
\citep{Bayer1999,Tischler2002}, we were able to extract $g_{e}^{z}$
and $g_{h}^{z}$ from the measured $g_{BE}^{z}$ and $g_{DE}^{z}$
\citep{Bayer2002,Gantz2016,Witek2011}. Eq.~\ref{eq: g_BE)DE)} is
derived from the parallel and anti-parallel spin nature of the DE
and BE, using the sign convention given by the Zeeman Hamiltonian Eq.~\ref{eq:Zeeman_Hamiltonian}.

If one takes into account also the excitonic fine structure due to the electron-hole exchange interaction, the BE and DE Zeeman splitting is given by \citep{Bayer2002}\textcolor{green}{}:
\begin{equation}
\Delta E_{BE(DE)}=\sqrt{\left(\delta_{1,2}^{1e1h}\right)^{2}+(\mu_{B}g^z_{BE(DE)}B_z)^{2}}\label{eq: Zeeman_square_root_fit}\ \ ,
\end{equation}
where $\delta_{1,2}^{1e1h}$ are the fine-structure splittings of the $X_{BE}^{0}$
and $X_{DE}^{0}$ at 0 field, respectively.
   
\begin{figure}
\centering{}\includegraphics[width=0.7\linewidth]{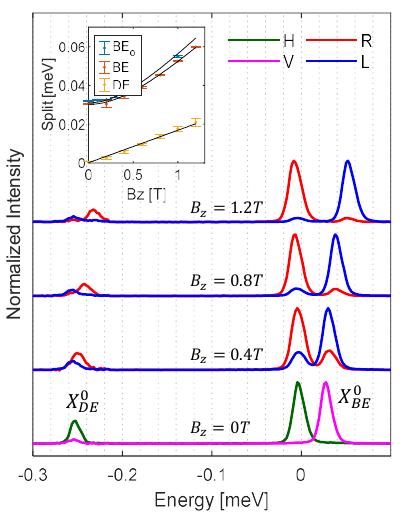}\caption[Zeeman splitting of the $X_{BE}^{0}$ and $X_{DE}^{0}$]{Polarization-sensitive magneto-PL of the bright and dark excitons ($X_{BE}^{0}$ and $X_{DE}^{0}$), for various magnetic field strengths.
The magnetic field has two components, one in the $\hat{z}$ direction which we vary, and one  which is fixed ($1.5T$) in $\hat{x}$ direction. The resulting Zeeman splittings are summarized in the Inset for both excitons using yellow (DE) and brown (BE)  data points. Blue data points in the Inset represent measurements of the BE splittings without the $\hat{x}$ field component. 
The errorbars represent the experimental uncertainties due to the finite spectral resolution of the measurements.
 Solid black lines represent best fits to Eq.~\Ref{eq: Zeeman_square_root_fit}, by which the $g$-factors
are extracted.
\label{fig:BE-DE-MPL}}
\end{figure}

In Figure \ref{fig:BE-DE-MPL}, we present polarization sensitive PL spectra of the QD neutral excitons  under various magnetic field strengths in Faraday configuration ($z$ direction).
In order to increase the visibility of the DE  we added during these measurements an in-plane magnetic field of 1.5 T.
In the inset to Figure \ref{fig:BE-DE-MPL} the measured Zeeman splittings of the DE and BE as a function of the $B_{z}$ field strength deduced from these measurements are displayed as yellow and brown data points, respectively, overlaid by solid black lines representing best fit of Eq~.\ref{eq: Zeeman_square_root_fit}. 
 In order to estimate the influence of the in-plane magnetic field on the measured Zeeman interaction we repeated the measurements of the BE in the absence of the in-plane field (blue data points and best fitted Eq.~\ref{eq: Zeeman_square_root_fit} in the inset to Figure~\ref{fig:BE-DE-MPL}).  The deduced $g$-factor of the BE with and without the in-plane field is  $-0.74$ and $-0.81$, respectively. 
This small difference (less than 10$\%$) sets an upper bound on the possible error in the measured $g$-factor of the DE. Moreover, we note that in the absence of $B_{z}$, the DE measured splittings is negligible, setting even a lower bound on the possible error in the value that we measure. This should not come as a surprise, since the electron and hole in-plane $g$-factors (which we measured in a separate experiment) are much smaller than those in the Faraday direction \citep{Witek2011}. 

One also notices in Figure \ref{fig:BE-DE-MPL} that the $X_{DE}^{0}$
cross-polarized doublet is not equally intense: at 0 field, its horizontally
(H) polarized component is much stronger than the vertically (V) polarized
one, a phenomenon observed and explained in previous publications
\textcolor{green}{}\citep{Schwartz2015b,Zielinski2015,Schmidgall2017}.\textcolor{green}{}
Adding magnetic field in Faraday configuration enhances the weaker
component and gradually increases the cross-circular polarization components of the 
the $X_{DE}^{0}$ doublet. However, up to the maximal field strength
of 1.5 T, the two $X_{DE}^{0}$'s components remain unequal. Nonetheless,
we extracted the $g$-factors of the $X_{BE}^{0}$ and $X_{DE}^{0}$
by fitting their measured Zeeman splittings to Eq.~\ref{eq: Zeeman_square_root_fit}.
 Table \ref{tab:g-factors} summarizes the values
of the measured excitonic $g$-factors, and the deduced single-carrier $g$-factors extracted from a simple arithmetic model, $g^z_{BE(DE)}=g^{z}_{1e}\pm g^{z}_{1h}$.

%\begin{figure}
%\centering{}\includegraphics[width=0.7\linewidth]{MPL_DE_BE}\caption[Zeeman splitting of the $X_{BE}^{0}$ and $X_{DE}^{0}$]{Polarization-sensitive magneto-PL of the bright and dark excitons ($X_{BE}^{0}$ %and $X_{DE}^{0}$), for various magnetic field strengths.
%The magnetic field has two components, one in the $\hat{z}$ direction which we vary, and one  which is fixed ($1.5T$) in the $\hat{x}$ direction. The resulting Zeeman splittings are summarized in the Inset for both %excitons using brown (DE) and blue (BE) data points. 
%For the BE, the datapoints are the values after substracting the $\hat{x}$ magnetic field component. 
%The errorbars represent the experimental uncertainties due to the finite spectral resolution of the measurements. Solid black lines represent best fits to Eq.~\Ref{eq: Zeeman_square_root_fit}, by which the $g$-%factors are extracted.
%\label{fig:BE-DE-MPL}}
%\end{figure}
 %\textcolor{blue}{If the data points in the inset are showing bars that are error bars, then NIST requires a statement of how the errors are calculated. Giora: Done.}

\begin{table}
\centering
\begin{tabular}{cccc}
\toprule 
Spectral line & $\delta_{1,2}^{1e1h}[\mu\text{eV}]$ & $g^{z}$-factor & Model\tabularnewline
\midrule
\toprule
$X_{BE}^{0}$ & $31(2)$ & $-0.81(2)$ & $g^{z}_{1e}+g^{z}_{1h}$\tabularnewline

$X_{DE}^{0}$ & $1.4(1)^{*}$ & $-0.29(4)$ & $g^{z}_{1e}-g^{z}_{1h}$\tabularnewline
\bottomrule
\midrule
%\end{tabular}

%\vspace{1em} % Add spacing if needed between the main table and the notes.

%\begin{tabular}{c}

\multicolumn{4}{l}{$ g_{1e}^{z}=-0.55(4)\quad ~~~ ;   \quad  ~~   \delta_{0}^{1e1h}=270(10)\mu eV$}\tabularnewline

\multicolumn{4}{l}{$g_{1h}^{z}=-0.26(4)\quad $}\tabularnewline

\end{tabular}

\caption[Excitonic lines parameters]{Measured excitonic fine structure splittings and $g$-factor tensor z-components.
Here $\delta_{1,2}^{1e1h}$ is the bright (dark) exciton line splitting at $B=0$, and $g_{(e/h)}^{z}$ are the $g$-factors of the electron and hole in Faraday configuration, respectively.\\
$^{*}$ The $X_{DE}^{0}$ zero field splittings $\delta_2^{1e1h}$ is too small to be directly resolved here. It was measured using
time-resolved spectroscopy in Ref.~\citep{Schwartz2015}. 
$\delta_{0}^{1e1h}$ denotes the measured $X_{BE}^{0}$ - $X_{DE}^{0}$ zero field splittings.}
\label{tab:g-factors}
\end{table}

\subsection{Diamagnetic shifts}

\begin{figure*}
\begin{centering}
\includegraphics[width=1\linewidth]{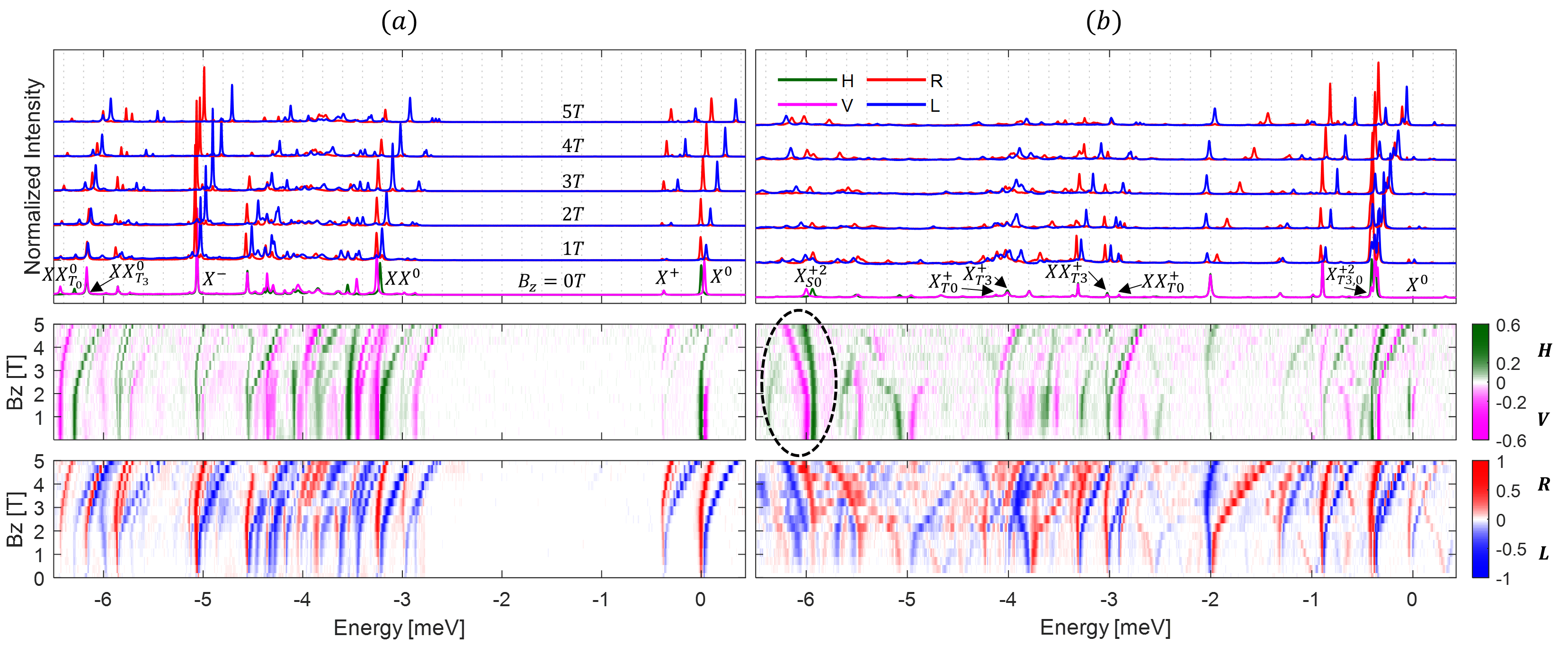}
\par\end{centering}
\caption[Full Magneto-PL.]{Polarization-sensitive magneto-PL spectra in Faraday configuration
for various magnetic field strengths, for negatively (a) and positively
(b) charged QD. The upper panel shows polarization-sensitive magneto-PL
spectra.  The panels below show the degree of circular and rectilinear
polarizations, given by the color bars to the right, as a function
of the photon energy and the externally applied magnetic field strength.
The identified spectral lines are marked: $X^{0}$ and $XX^{0}$- the neutral bright exciton and biexciton,
and $XX_{T_{0}(T_{3})}^{0}$ the neutral metastable biexcitons with
the two holes in $T_{0}$($T_{3})$ spin triplet configurations. The $X^{+}$, ($X^{-}$) positively (negatively)
charged trion,  $X_{T_{0}(T_{3})}^{+}$,
and $XX_{T_{0}(T_{3})}^{+}$ are similar positively charged excitons
and biexcitons. The $X^{+2}$ lines result from the recombination
of the doubly positively charged exciton, leaving behind two holes
that can form either a singlet $S_{0}$ or one of the triplets, $T_{\pm3}$
or $T_{0}.$  Note the negative diamagnetic shift of the $X_{S_{0}}^{+2}$
(marked with an oval dash line). The energy scale is relative to the
$X_{BE}^{0}$ spectral line at zero magnetic field. 
The $X_{BE}^{0}$ spectral line is clearly observed when the QD is statistically closed to neutrality (a), but the line is suppressed when the QD is strongly positively charged (b). \label{fig:MPL}}
\end{figure*}

In Figure \ref{fig:MPL}, we present a full set of Magneto-PL measurements
in Faraday configuration of the QD.
We present the PL spectra for two average charge states of the QD: negative and
positive. The QD average charge state is apparent in each case by considering
the emission ratio between the positive and negative trions, $X^{+}$
and $X^{-}$. Many identified lines are marked in the PL following
previous studies \citep{Benny2012}.

All the observed spectral lines split into two components under the application of external magnetic field along the optical axis $z$. 
At high fields the two components are mostly cross-circularly polarized as can be seen in the lowest  panel of Figure \ref{fig:MPL}. 
The line splitting is known to be due to the Zeeman interaction between the confined carriers and the external magnetic field as expressed in Eq.~\ref
{eq:Zeeman_Hamiltonian}

On top of the  splitting due to the Zeeman interaction, the spectral lines undergo a clear quadratic-in-$B$ (diamagnetic) shift. 
This shift is well described for all the measured magneto photo-luminescence lines by adding a term 
 $\beta B^{2}$   to the Hamiltonian (Eq.~\ref{eq:Zeeman_Hamiltonian}). 
This term faithfully describes the experimentally measured shifts for all the spectral lines as shown in Figure
\ref{fig:MPL}. The diamagnetic shift refers to the spectral "center
of mass", defined as  $E(B)=(E_{R}(B)+E_{L}(B))/2$, where $E_{R}$ and $E_{L}$ are the energies of the two Zeeman components emitting in cross circular $R$ and $L$ polarizations, respectively.
In most cases, $E(B)$ shifts towards higher energy (hence the terminology of ``diamagnetic'' versus ``paramagnetic'' shift). 

\begin{figure}
\centering{}\includegraphics[width=1\linewidth]{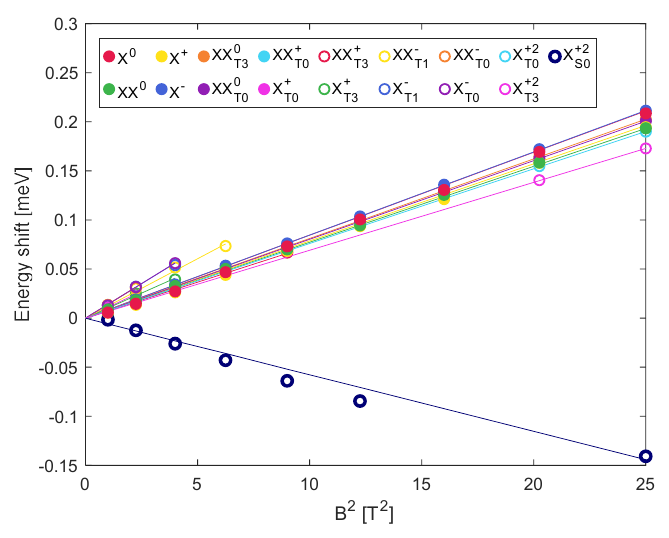}\caption[Diamagnetic shifts summary versus $B^{2}$]{Measured energy shifts of various optical transitions as a function
of $B^{2}$.  The $X_{S_{0}}^{+2}$ spectral  line is a prominent exception - it shows a negative diamagnetic shift.
\label{fig:dia_summary} }
\end{figure}
Figure \ref{fig:dia_summary} summarizes the diamagnetic shifts of
several selected lines. One can see that many lines, including the
$XX^{0}$ and the trions, $X^{-}$ and $X^{+}$, exhibit
similar diamagnetic shifts to that of the BE ($\sim8~\mu eV/T^{2}$).

%In section IV, below,  we compare all the measured diamagnetic shifts with our model calculations.

\subsection{The diamagnetic shifts of the $X^{+2}$ spectral lines} 

\begin{figure}
\centering{}\includegraphics[width=1\linewidth]{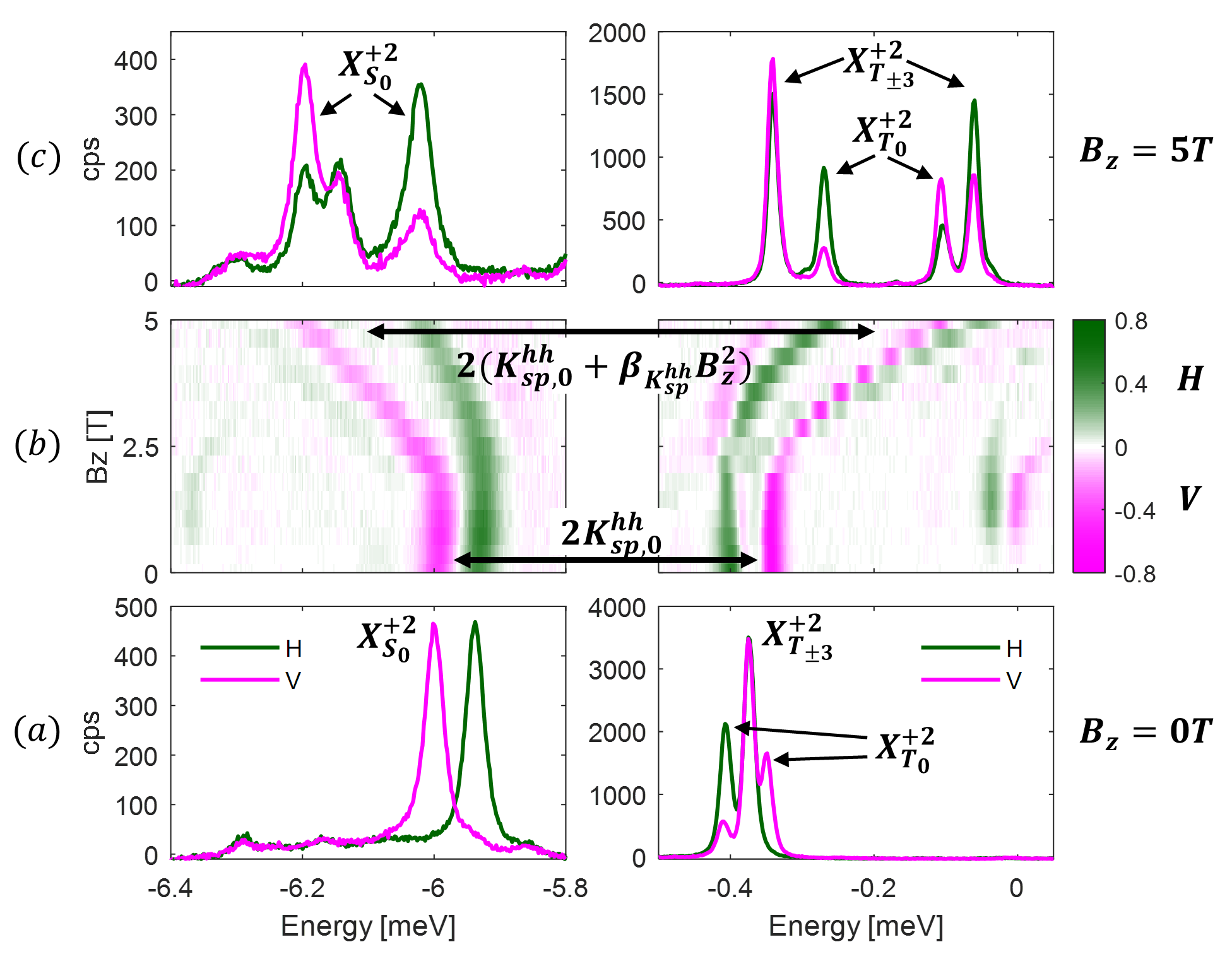}\caption[Magneto-PL of $X_{S_{0}}^{+2}$ and $X_{T_{0}}^{+2}$]{Rectilinear polarization-sensitive PL spectra of the $X^{+2}$ spectral
lines relative to the neutral exciton state a) at zero magnetic field,
b) as function of the externally applied field in Faraday configuration,
and c) in magnetic field of $5T$. The transitions are marked by their
final spin configurations ($S_{0},$$T_{0}$, $T_{\pm3}$). The energy
difference between the $X_{T_{0}}^{+2}$ and the $X_{S_{0}}^{+2}$
doublets (marked) equals twice the hole-hole exchange interaction $K_{sp,0}^{hh}$
\label{fig:Xp2-MPL}}
\end{figure}

Interestingly, a few spectral lines in Figure \ref{fig:MPL}(b) when the QD is positively charged shift towards lower energy as the external field increases.  In particular, one prominent spectral line at $-5.95~\text{meV}$ (at zero field) exhibits such a distinctive
\textit{negative} diamagnetic shift. We focus our discussion on this particular line (represented by the blue circles in Figure \ref{fig:dia_summary}). 
We identify this line as the doubly charged exciton transition $X_{S_{0}}^{+2}$, where the subscript ${S_{0}}$ refers to the singlet configuration of the two holes in the final state of this transition. The other triplet configurations of the final states result in the 3 optical transitions at about $-0.38~\text{meV}$. We denote them by $X_{T_{0}}^{+2}$ and $X_{T_{\pm3}}^{+2}$. 
These spectral lines are presented on an expanded energy scale in Figure \ref{fig:Xp2-MPL}.
The other less prominent lines which also exhibit negative shifts result from 2 holes singlet states in which the second hole is in a higher level than the first hole. For the sake of simplicity, they will not be further discussed in this work.

\begin{figure}
\centering{}\includegraphics[width=0.95\linewidth]{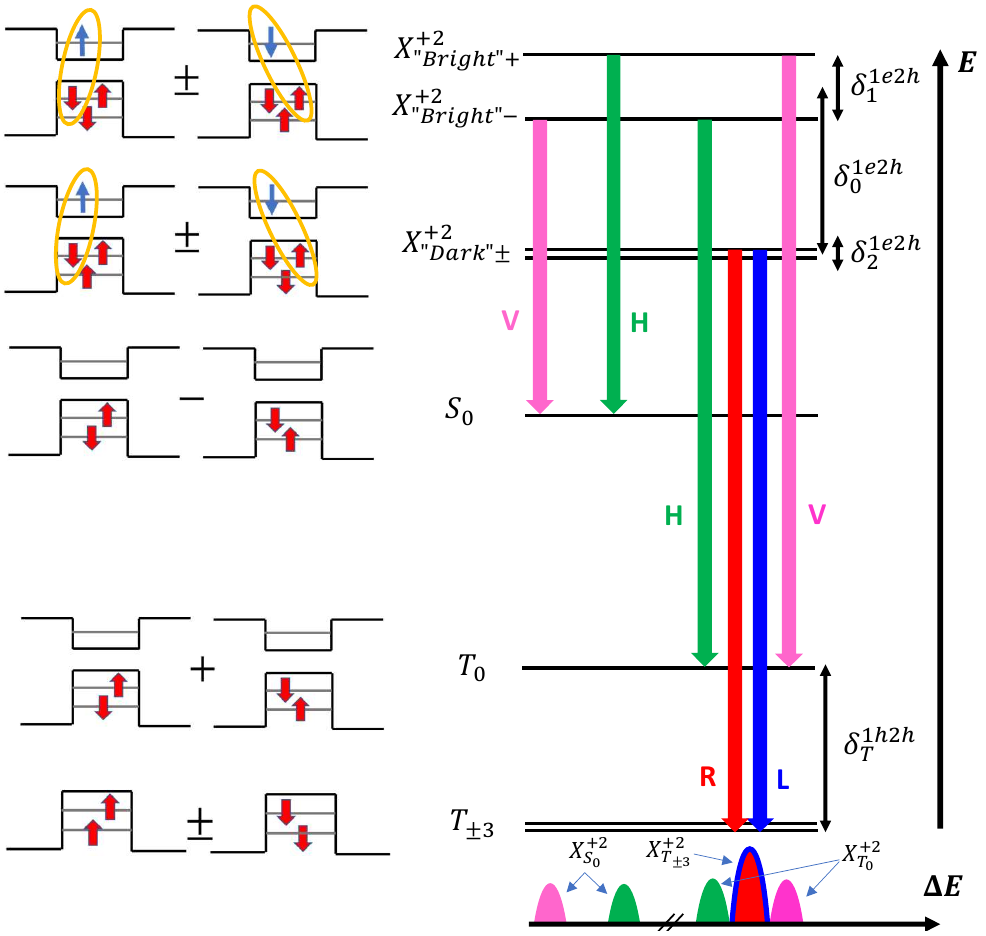}\caption[$X^{+2}$ energy level scheme and optical selection rules]{Schematic description of the energy levels and optical 
transitions associated with the doubly positively charged exciton $X^{+2}$. The
configuration of each state is presented on the left, where thin blue
arrows represent electrons with spin $\frac{1}{2}$, and thick arrows
represent heavy-holes with spin $\frac{3}{2}$. The polarization selection
rules are marked by colored downward arrows. $H$ ($V$) marks the
horizontal (vertical) rectilinear polarization, while $R$ ($L$) marks the circular right- (left-) handed polarization. A schematic description
of the emitted PL is drawn at the bottom. The $X_{T_{\pm3}}^{+2}$
spectral line is drawn in red with a blue edge, symbolizing that
the $R$ and $L$ polarizations overlap such that the emission is
unpolarized. \label{fig:Xp2-diagram} }
\end{figure}

The identification of the spectral lines follows Ref. \citep{Benny2012}, where the same QD was studied. The identifications are based on the following arguments. (i) these spectral lines appear in the spectrum only when the QD is strongly positively charged, like in Figure \ref{fig:MPL}b. (ii) the $X_{S_{0}}^{+2}$ and $X_{T_{0}}^{+2}$ transitions exhibit the same fine-structure splitting of $\sim 70~\mu\text{eV}$ due to the same splitting in their initial state, denoted $\delta_{1}^{1e2h}$ (see detailed energy level scheme in Figure \ref{fig:Xp2-diagram}). (iii) The energy difference between these states is $\sim5~\text{meV}$, matching previously measured hole-hole exchange interaction energies, for example, between the excited positive trion states $(1e^{1})(1h^{1}2h^{1})_{S}$ and $(1e^{1})(1h^{1}2h^{1})_{T}$, or between the positive biexciton states $(1e^{2})(1h^{2}2h^{1})$ \citep{Benny2012}.  
Here, $np^{m}$ reads: $n$ - the energy level
order, $p$ - the particle type ($e$ or $h$), and $m$ - the number
of particles occupying this level (either 1 or 2). The subscript of parentheses that includes two carriers of the same type, describes their mutual spin configuration - either a singlet, $S_0$, or triplet, $T_l$, where $l=0,\pm1$ for two electrons or 0, $\pm3$ for two heavy-holes, denotes the total spin projection on the $z$ axis.   

Figure \ref{fig:Xp2-diagram} schematically describes the energy levels
and the optical transitions associated with the doubly charged exciton,
$X^{+2}$. This exciton comprises one electron in the ground-level
$1e^{1}$, and three holes: two of them forming a singlet in the s-shell
ground-level, $1h^{2}$, and the third one occupies the first excited
p-shell level $2h^{1}$.  The exchange interaction
between the Pauli-unpaired electron (in the 1st level) and the Pauli-unpaired hole (in the
2nd level) removes the degeneracy between the four possible two-carriers'
spin configurations, forming four distinct eigenstates similar to
the case of the neutral exciton ($X^{0}$). As such, we borrow the
exciton ``bright'' and ``dark'' terminology to describe the eigenstates
of the $X^{+2}$ as well. States with anti-parallel e-h spins are
called ``bright-like'', while states with parallel spins - ``dark-like''
(see Figure \ref{fig:Xp2-diagram}). We emphasize that the dark and
bright states are both optically active since the optical recombination
occurs mostly between the unpaired $s-$electron and one of the $s-$level
singlet holes, rather than with the unpaired $p-$hole. The $1e^1$  -  $2h^1$ recombination is very inefficient (about two orders of magnitude weaker~\citep{Schwartz2015}) since it is forbidden by symmetry.

The final states of the $X^{+2}$ recombination contain two holes
- one in the ground level and one in the first excited level. As identical
particles, they form either one singlet spin state denoted by $S_{0}^{1h2h}$
or three triplet states denoted by $\left\{ T_{0}^{1h2h},T_{\pm3}^{1h2h}\right\} $,
respectively. The two initial bright-like exciton states can only
recombine to the singlet $S_{0}^{1h2h}$ or triplet $T_{0}^{1h2h}$
final states (but not to the $T_{\pm3}^{1h2h}$), resulting in two
pairs of cross-rectilinearly polarized doublets \citep{Kodriano2010};
the dark-like states can only recombine to the $T_{\pm3}^{1h2h}$
states. Because in the absence of external magnetic field the dark-like states and the $T_{\pm3}^{1h2h}$ states are each nearly degenerate,
the recombination results in a single, unpolarized, strong spectral
line. We label the $X^{+2}$ optical transitions by their final states,
specified in their subscripts: $X_{T_{0}}^{+2}$, $X_{T_{\pm3}}^{+2}$
and $X_{S_{0}}^{+2}$. The latter transition, $X_{S_{0}}^{+2}$, is
the one exhibiting a negative diamagnetic shift. We note that in the
absence of external field, the unpolarized $X_{T_{\pm3}}^{+2}$ spectral
line is positioned exactly in between the two cross linearly polarized
components of the $X_{T_{0}}^{+2}$ line. This indicates that $\delta_{0}^{1e2h}$,
denoting the splitting between the dark-like and bright-like $X^{+2}$
states, is equal to $\delta_{T}^{1h2h}$, the splitting between the
holes' triplet states $T_{0}^{1h2h}$ and $T_{\pm3}^{1h2h}$\textcolor{green}{}.

A detailed polarization-sensitive magneto-PL spectra of the $X^{+2}$
spectral lines are presented in Figure \ref{fig:Xp2-MPL}. One can
see that while the triplet lines shift towards higher energy with
increasing B-field, the singlet lines shift towards lower energy.
Since the initial states of the $X_{S_{0}}^{+2}$ and $X_{T_{0}}^{+2}$
transitions are the same (the bright-like exciton states), we conclude
that the sign difference between the diamagnetic shifts of the two transitions stems from the different influence that the external
magnetic field has on the final states. The h-h singlet final state
rises in energy faster than the initial state such that the
overall spectral shift is negative (red shift). On the other hand, the h-h triplet
state rises in energy slower than the initial state, thus the
total spectral shift is positive (blue shift).

\subsection{The measured $g$-factors of the charged excitons $X^{\pm1}$ and $X^{+2}$ \label{subsec: X+2 g-factor}}
The measured $g$-factors of the charged excitons $X^{\pm}$, and $X^{+2}$
optical transitions are summarized in Table \ref{tab:xp2-g-factors} together with that of the neutral bright exciton $X^{0}$. 
In discussing these $g$-factors, one should consider the Zeeman interaction both in the initial and in the final states of the optical transitions. For example, 
in the initial state of the $X^{\pm}$ exciton the unpaired electron (hole) interacts with the field while in the final state the remaining hole (electron) interacts with the field. Thus, like in the case of the BE, one expects that the measured spectral line $g$-factor is a sum of the electron and hole s-level $g$-factors. 
Similarly, the final states of the bright-like $X^{+2}$ transitions are either the two-holes $T_0$ triplet or $S_0$ singlet states. Both states have zero $g$-factors, since they are non-degenerate and their total angular momentum projection on the magnetic field direction vanishes. Therefore, the $g$-factors of these transitions are due to the initial bright-like exciton state only. This state contains a Pauli-unpaired electron in the first s-level and a Pauli-unpaired hole in the second p-level. The two paired holes occupying the ground level singlet state do not contribute to the Zeeman interaction with an externally applied magnetic field. Thus, the measured $g$-factors of the $X^{+2}_{T_0}$ and $X^{+2}_{S_0}$ lines are expected to be the same. In addition, like the BE, their zero field splitting is given by the anisotropic electron-hole exchange term $\delta_1^{1e2h}$. 
%and possible magnetic field dependence of the anisotropic exchange interaction ($\delta_{1}^{1e2h}$) between the Pauli unpaired electron and hole. 
The overall $g$-factor of this transition is therefore:
\begin{equation}
(E^{initial}-E^{final})/(\mu_{B}B_{z})=g_{1e}^{z}-(-g_{2h}^{z})=g_{1e}^{z}+g_{2h}^{z}\label{eq:g_aritmetics}
\end{equation}
The extra minus sign results from the optical selection rule stating that only an electron and a hole with anti-parallel spins can radiatively recombine.

The lines $g$-factors are extracted by fitting their measured Zeeman splittings to Eq.~\ref{eq: Zeeman_square_root_fit}.
 Table \ref{tab:xp2-g-factors} summarizes the extracted values. 
The results of this simple arithmetic calculation are presented in the 'Model' column of  Table \ref{tab:xp2-g-factors}.
Comparing this simple model with experiments, one indeed finds that the measured $g$-factors of spectral lines like $X^{+}$, $X^{-}$, 
and other optical transitions whose $g$-factor arithmetics produces the sum of the electron and hole ground level $g$-factors, are similar to the measured value of the BE,  \citep{MscGiora,Ayal_Thesis}. The same arithmetic shows that $g$-factor of the doubly-charged exciton triplet states $X^{+2}_{T_{\pm3}}$ is yet another example. 
However, the $g$-factors of the $X^{+2}_{S_{0}}$ and $X^{+2}_{T_{0}}$ transitions are similar to those that include the $g$-factor 
of the hole in its second energy level ($g^z_{2h}$) rather then the first ($g^z_{1h}$) \citep{MscGiora,Ayal_Thesis}.  
It is interesting to note that it follows (see Table \ref{tab:xp2-g-factors}) that $g_{2h}^{z}=1.15\pm0.10$. 
This value is opposite in sign compared to the ground level $g$-factors of the hole and that of the
electron ($-0.26$ and $-0.55$, respectively (see Table \ref{tab:g-factors}). 
The opposite sign of the excited-hole $g$-factor was directly measured on another similar QD sample using PLE spectroscopy \citep{Ayal_Thesis}. 
This finding is also supported by realistic NextNano\textsuperscript{TM}\cite{NISTdisclaimer} simulations \citep{MscGiora}. 
Similar results of sign difference between the first and second confined levels were 
observed and discussed for electrons and holes in quantum wells~\cite{Ivchenko2005,Kiselev1998}.

\textcolor{red}{ }

\begin{table}
\begin{centering}
\begin{tabular}{clcc}
\toprule 
Spectral line & $ZFS[\mu eV]$ & $g$-factor  & Model\tabularnewline
\bottomrule 

\midrule 
$X_{BE}^{0}$ & $31(2)$ & $-0.81(2)$  & \multirow{4}{*}{$g_{1e}^{z}+g_{1h}^{z}$}\tabularnewline
%\cmidrule{1-3} \cmidrule{2-3} %\cmidrule{3-3} 
$X^{+}$ & $0$ & $-0.81(2)$ & \tabularnewline
%\cmidrule{1-3} \cmidrule{2-3} %\cmidrule{3-3}
$X^{-}$ & $0$ & $-0.94(2)$ & \tabularnewline
%\cmidrule{1-3} \cmidrule{2-3} %\cmidrule{3-3}  
$X_{T_{\pm3}}^{+2}$ & $0$ & $-0.97(5)$ & \tabularnewline

\midrule 
$X_{T_{0}}^{+2}$ & $56(2)$ & $0.55(10)$ & \multirow{2}{*}{$g_{1e}^{z}+g_{2h}^{z}$}\tabularnewline
%\cmidrule{1-3} \cmidrule{2-3} \cmidrule{3-3} 
$X_{S_{0}}^{+2}$ & $69(7)$ & $0.65(7)$ & \tabularnewline
\midrule 
\bottomrule
\end{tabular}
\par\end{centering}
%\centering{}
\begin{tabular}{c}
%\toprule 

$g_{2h}^{z}=1.15(12)\quad  ;\quad \delta_{0}^{1e2h}=130(10)~\mu eV$\tabularnewline

$$\tabularnewline
\end{tabular}
\caption{ $g$-factors and zero field splitting (ZFS) of the charged excitons spectral lines transitions,
compared to those of the neutral excitons. The measured splittings are explained
by a simple model which assumes that the $g$-factor of a given transition
can be decomposed to the sum of the comprising charge carrier $g$-factors
of the initial and final states of the optical transition. $g_{n(e/h)}^{z}$
denotes the $g$-factor of the electron (hole) in the $n$ energy
level of the QD, where $n=1$ is the ground level.  \label{tab:xp2-g-factors}}
\end{table}

\section{Theoretical Analysis}
\label{sec:Theoretical Analysis}

\subsection {The Hartree - Fock approximation applied to QD-confined multi-carrier configurations} 
We use the  Hartree-Fock approximation \citep{Roothaan1951} for describing the multi-carrier configurations involved in the various initial and final states of the optical 
transitions that we study. In this approximation, the many-body state is a single Slater determinant of single-particle states. The single-particle states are found as confined states in the dot, described by a parabolic model-potential, in the strong-confinement limit where any possible single-particle reshaping by interaction with the other particles is ignored.  One expresses the energies of the many-body states as sums of single-particle energies, plus the direct Coulomb and exchange interactions between all pairs of particles in the particular configuration.  

For example, the energy of the neutral bright (BE) and dark (DE) excitons are given by:
\begin{equation}
E_{X^{0}_{BE}}=E_{s}^{e}+E_{s}^{h}-J_{ss}^{eh},\label{eq:E_X0BE}
\end{equation}
and 
\begin{equation}
E_{X^{0}_{DE}}=E_{s}^{e}+E_{s}^{h}-J_{ss}^{eh}-2K_{ss}^{eh}=E_{X^{0}_{BE}}-2K_{ss}^{eh},\label{eq:E_X0DE}
\end{equation}
where $E_{s}^{e\left(h\right)}$
is the confinement energy of the electron (hole) in the $s$-level and $J_{ss}^{eh}$ ($K_{ss}^{eh}=0.5 \delta_0^{1e1h}$)
is the direct Coulomb (exchange) interaction between the electron and the hole. We note here that as a rule, one subtracts twice the exchange energy when the spins of the two particles are aligned \citep{Roothaan1951}.  
%This is the case for the two holes in the triplet state $T^{1h2h}_{\pm 3}$ \citep{Roothaan1951}.  
This rule  applies also for the  case of electron and hole pair though they are distinguishable quasi-particles \citep{Takagahara2000,Hawrylak}.
The electron-hole anisotropic exchange interactions, $\delta_1^{1e1h}$  ($\delta_2^{1e1h}$) which removes the degeneracy between the bright (dark) exciton
eigenstates is usually much smaller than $\delta_0^{1e1h}$ \cite{Poem2007}, and does not affect the excitonic transitions' spectral center of gravity. Therefore, we ignore these, as  well as  $\delta_1^{1e2h}$  ($\delta_2^{1e2h}$) in this discussion.  

Applying the same rules to the positively charged exciton, the $X^{+1}$, in which the initial state is formed from two holes in their respective $s$-level and one electron is in its respective $s$-level, one gets,    
\begin{equation}
E_{X^{+1}}^{init}=E_{s}^{e}+2E_{s}^{h}-2J_{ss}^{eh}-2K_{ss}^{eh}+J_{ss}^{hh}.\label{eq:E_X+init}
\end{equation}
After the electron-hole recombination, the final state contains only one hole in its $s$ level
 \begin{equation}
 E_{X^{+1}}^{final}=E_{s}^{h}.\label{eq:E_X+final}
 \end{equation}
 The optical transition energy is therefore,
 \begin{eqnarray}
 E_{X^{+1}}=E_{X^{+1}}^{init}-E_{X^{+1}}^{final}&=&E_{s}^{e}+E_{s}^{h}-2J_{ss}^{eh}-2K_{ss}^{eh}+J_{ss}^{hh}\nonumber\\
 &=&E_{X^{0}_{DE}}+J_{ss}^{hh}-J_{ss}^{eh}.\label{eq:E_X+}
 \end{eqnarray} 
 
 It turns out, that the energy difference between the $X^{+1}$ and the $X^{0}_{DE}$ transitions is given by the difference between the two direct Coulomb terms on the right hand side of Eq.~\ref{eq:E_X+}.  

We proceed by applying the same rules to the doubly positively charged exciton, the $X^{+2}$ spectral lines. Considering first the energies of the initial states,
\begin{eqnarray}
E_{X_{S_{0}}^{+2}}^{initial}=E_{X_{T_{0}}^{+2}}^{initial}&=&E_{s}^{e}+2E_{s}^{h}+E_{p}^{h}+J_{ss}^{hh}+2J_{sp}^{hh}\nonumber\\
& &-2J_{ss}^{eh}-J_{sp}^{eh}-2K_{ss}^{eh}-2K_{sp}^{hh}\label{eq:HF_init}
\end{eqnarray}
\begin{eqnarray}
%E_{X_{T_{\pm3}}^{+2}}^{initial}&=&E_{s}^{e}+2E_{s}^{h}+E_{p}^{h}+J_{ss}^{hh}+2J_{sp}^{hh}\nonumber\\
%& &-2J_{ss}^{eh}-J_{sp}^{eh}-2K_{sp}^{hh}-2K_{ss}^{eh}-2K_{sp}^{eh},\label{eq:HF_init}
E_{X_{T_{\pm3}}^{+2}}^{initial}&=&E_{X_{T_{0}}^{+2}}^{initial}-2K_{sp}^{eh},\label{eq:HF_init}
\end{eqnarray}

and then of the final states, 
\begin{eqnarray}
E_{X_{S_{0}}^{+2}}^{final}&=&E_{s}^{h}+E_{p}^{h}+J_{sp}^{hh}\label{eq:HF_final}.
\end{eqnarray}
\begin{eqnarray}
E_{X_{T_{0}}^{+2}}^{final}=E_{X_{T_{\pm3}}^{+2}}^{final}&=&E_{s}^{h}+E_{p}^{h}+J_{sp}^{hh}-2K_{sp}^{hh}\label{eq:HF_final}.
\end{eqnarray}
We note that here as well, we subtract twice the exchange energy whenever a pair of particles has parallel spins. 

The transition energies are then given as before by  the differences between the initial and final states,   
\begin{eqnarray}
E_{X_{T_{0}}^{+2}}=E_{X^{+1}}+J_{sp}^{hh}-J_{sp}^{eh},\label{eq:T0} 
\end{eqnarray}
\begin{eqnarray}
E_{X_{S_{0}}^{+2}}=E_{X_{T_{0}}^{+2}}-2K_{sp}^{hh},\label{eq:S0} 
\end{eqnarray}
\begin{eqnarray}
%E_{X_{T_{\pm3}}^{+2}}=E_{X^{+1}}+J_{sp}^{hh}-J_{sp}^{eh}-2K_{sp}^{eh}=E_{X_{T_{0}}^{+2}}-2K_{sp}^{eh},\label{eq:T3}
E_{X_{T_{\pm3}}^{+2}}=E_{X_{T_{0}}^{+2}}-2K_{sp}^{eh},\label{eq:T3}  
\end{eqnarray}
where we also used Eq.~\ref{eq:E_X+}.
It follows that the energy difference between the $X^{+1}$ spectral line and the $X_{T_{0}}^{+2}$ spectral line is given by the difference between the two direct Coulomb terms  $J_{sp}^{hh}$ and $J_{sp}^{eh}$. 

\subsection{The 2D parabolic potential model}

In order to quantitatively compare our model (Eqs.~\ref{eq:E_X0DE}, \ref {eq:E_X+} ,\ref{eq:T0}, \ref{eq:S0} and \ref{eq:T3}) with the measured PL spectra, we proceed with a model for calculating the direct Coulomb and exchange interactions. We follow Warburton et. al.~\citep{Warburton1998} who developed a simple two-dimensional, cylindrically symmetric, harmonic oscillator model for describing the QD confining potential for charge carriers. 
In their model, the harmonic potential is described by two parameters: $\omega_{e(h)}$, the harmonic frequency of
the confining potential and $m_{e(h)}$, the in-plane effective mass, for each one of the confined carriers, the electron ($e$) or the heavy hole ($h$), respectively.
The effective length of the potential $l_p$ is related to the effective mass and frequency by 
\begin{equation}
l_{p}=\sqrt{\frac{\hbar}{m_{p}\omega_{p}}}.\label{eq: Harmonic_length1}
\end{equation}
where $l_{p}$ is the effective length for particle type $p\in\{e,h\}$.

The single particle wavefunctions for the two lowest energy levels in this model are:   
\begin{equation}
\Psi^{1p}(\rho)=\frac{1}{\sqrt{\pi}l_p}{e^{-\frac{\rho^2}{2l_p^2}}}; \Psi^{2p}_{\pm1}(\rho,\phi)=\frac{\rho}{\sqrt{\pi}l_p^2}{e^{(-\frac{\rho^2}{2l_p^2}\pm i\phi)}}\label{eq:WF1} 
\end{equation}
where $\rho$  and $\phi$ are the conventional inplane particle's radius vector length and polar angle, respectively.  
 
The various direct Coulomb interaction integrals, $J^{p_ip_j}_{ij}$, where $i,j$ indicate the confined carriers' states and $p_i,p_j$ indicate their types, can be thus analytically calculated \citep{Warburton1998}. 
Relevant to our discussion, as discussed above, are the direct and exchange terms between holes and electrons in their respective lowest two energy levels ``$s$- and $p$-shells''. 
Interactions between particles of same type are inversely proportional to $l_p$ \citep{Warburton1998} 
\begin{equation}
J_{ss}^{pp}=\frac{4}{3}J_{sp}^{pp}=4K_{sp}^{pp}=\frac{\alpha}{l_p},\label{eq:JE} 
\end{equation}
with
\begin{equation}
\alpha=\sqrt{\frac{\pi}{2}}\frac{e^{2}}{4\pi\epsilon_{0}\epsilon_{r}}.\label{eq:alpha}
\end{equation}
Here $e$ is the electron charge, $\epsilon_{0}$ is the vacuum permittivity
and $\epsilon_{r}$ is the relative permittivity (dielectric constant) of the QD material.

In a similar way the direct Coulomb interactions between carriers of different types are given by: 
\begin{equation}
J_{ss}^{pp'}=\alpha\left(\frac{2}{l_p^2+l_{p'}^2}\right)^{\frac{1}{2}}; J_{sp}^{pp'}=\frac{\alpha}{\sqrt{2}}\frac{2l_p^2+l_{p'}^2}{\left(l_p^2+l_{p'}^2\right)^{\frac{3}{2}}}.\label{eq:JEHSSSP} 
\end{equation}

Since the electron and the hole are distinguishable quasi-particles \citep{Takagahara2000,Hawrylak} the exchange terms, $K_{ss}^{eh}$ and $K_{sp}^{eh}$ are not directly given by the Hartree Fock model. 
Yet, since these exchange terms are known to be mainly short range they can be approximated by\citep{Takagahara2000,Ivchenko2005}   
\begin{equation}
K_{ij}^{eh}\approx E^{eh}_{sr}a_0^3\int\left|f^{X}_{ij}(\mathbf{r},\mathbf{r})\right|^2d\mathbf{r}\label{eq:KEHR}
\end{equation}
where $E^{eh}_{sr}$ is the short range electron-hole exchange interaction, $a_0$ is the crystal lattice constant\citep{Takagahara2000, Ivchenko2005} and  $f^X_{i,j}(\mathbf{r_e, r_h})$ is the envelope function of the exciton formed by electron in the i level and hole in the j level. We proceed by approximating the exciton envelope functions as a product of the single carrier envelope functions $f^X_{i,j}(\mathbf{r_e, r_h})\approx \Psi^{ie}(\boldsymbol{r_e})\Psi^{jh}(\boldsymbol{r_h})$. This approximation is valid for QDs in which the confining potential is larger than the electron hole direct Coulomb interaction. Hence:

\begin{equation}
K_{ss}^{eh}\approx E^{eh}_{sr}a_0^2\int\left|\Psi^{1e}(\boldsymbol{\rho})\right|^2\left|\Psi^{1h}(\boldsymbol{\rho})\right|^2d\boldsymbol{\rho}=\frac{E^{eh}_{sr}a_0^2}{l_e^2+l_h^2} \label{eq:KEHR1}
\end{equation}
\begin{equation}
K_{sp}^{eh}\approx E^{eh}_{sr}a_0^2\int\left|\Psi^{1e}(\boldsymbol{\rho})\right|^2\left|\Psi^{2h}(\boldsymbol{\rho})\right|^2d\boldsymbol{\rho}=\frac{E^{eh}_{sr}a_0^2l_e^2}{(l_e^2+l_h^2)^2} \label{eq:KEHR2}
\end{equation}
where we used in Eqs. \ref{eq:KEHR1} and \ref{eq:KEHR2} the corresponding single carrier envelope functions from Eq. \ref{eq:WF1}. 

The ratio between these exchange terms is therefore given by:
\begin{equation}
\frac{K_{sp}^{eh}}{K_{ss}^{eh}}=\frac{l_e^2}{l_e^2+l_h^2}=\frac{\gamma^2}{1+\gamma^2}\label{eq:KEHR3}
\end{equation}
where we define the ratio between the characteristic lengths of the carriers' potential: 
 $\gamma=\tfrac{l_e}{l_{h}}$. 

Experimentally, as we show below for the QD under study,  it turns out that $l_e$ and $l_h$ are nearly equal and  $\gamma \approx 1$.
Therefore by substituting $\gamma=1+\eta$, with $\abs \eta \ll 1$ in Eqs.~\ref{eq:KEHR3} and keeping only leading terms in $\eta$, one gets that   
\begin{equation}
 \frac{K_{sp}^{eh}}{K_{ss}^{eh}}\approx\frac{1}{2}(1+\eta)\approx\frac{1}{2}\label{eq:KEHR4}
\end{equation}

In a similar way one gets for the direct Coulomb and same particle exchange terms that:
\begin{equation}
J_{ss}^{hh}-J_{ss}^{eh}=4(J_{sp}^{hh}-J_{sp}^{eh})\approx \frac{1}{2}\eta J_{ss}^{hh}=2\eta K_{sp}^{hh}.\label{eq:eta1} 
\end{equation}

In addition, it is worth noting that the direct Coulomb attraction between the electron and the hole in their respective lowest energy levels $J_{ss}^{eh}$ can be used as 
an estimate for the BE binding energy $E_{BE}$. Thus using Eq.~\ref{eq:eta1} and Eq.~\ref{eq:JE} we get:
 \begin{equation}
E_{BE}\approx J_{ss}^{eh}\approx J_{ss}^{hh}(1-\frac{1}{2}\eta)=4K_{sp}^{hh}(1-\frac{1}{2}\eta).\label{eq:E_BE} 
\end{equation}
 Eqs.~\ref{eq:KEHR3}, \ref{eq:eta1} and \ref{eq:E_BE} are useful since, as we show below, the hole-hole and electron-hole exchange energies $K_{sp}^{hh}$, $K_{sp}^{eh}$ and $\eta$ can be determined experimentally from the PL spectra at zero external magnetic field. 
 
 \subsection{The magnetic field dependence of the direct and exchange Coulomb interactions} 
 
To include the effect of the magnetic field in this model,
we replace $\omega_{p}$ with $\omega_{p}(B)\equiv\sqrt{\omega_{p}^{2}+\frac{e^{2}B_{z}^{2}}{4m_p^{2}}}$,
obtained by adding a magnetic field Hamiltonian to the harmonic oscillator
one and solving for the eigenenergies (harmonic spectrum plus Landau
levels spectrum). Using Eq.~\ref{eq: Harmonic_length1} the expression for the effective length then becomes
\begin{equation}
l_{p}(B)%=l_{p,0}\left[1+\left(\frac{eB}{2m_p\omega_{p}}\right)^2\right]^{-\frac{1}{4}}
= l_{p,0}\left[1+\left(\frac{el_{p,0}^2}{2\hbar}\right)^2 B^2\right]^{-\frac{1}{4}} \label{eq:eff_length_B},
\end{equation}
where we add the subscript ``0'' (such as in $l_{p,0}$) to indicate the value in the absence of magnetic field.

In low magnetic fields, the magnetic energy is much smaller than the confinement energy, such that 
%$\left(\frac{eB}{2m\omega_{p}}\right)^{2}=%
$\left(\frac{el_{p,0}^2}{2\hbar}\right)^2B^2\ll1$, and to lowest terms in B:
\begin{equation}
l_p(B)\approx l_{p,0}\left[1-\left(\frac{el_{p,0}^2}{4\hbar}\right)^2 B^2\right]. \label{eq:eff_length_Bapp}
\end{equation}
Since by Eq.~\ref{eq:JE}, $K_{sp,0}^{pp}$ is inversely proportional to $l_{p,0}$, it follows from Eq.~\ref{eq:eff_length_Bapp} that the field dependence of the exchange energy can be approximated by 
%\begin{equation}
%K_{sp}^{hh}(B)\approx K_{sp}^{hh}\left[1+\left(\frac{e\alpha^2}{4\hbar{K_{sp}^{hh}}^2}\right)^2 B^2\right]\label{eq:E_sp-1}
%\end{equation}
\begin{equation}
K_{sp}^{pp}(B)\approx K_{sp,0}^{pp}+\beta_{K_{sp}^{pp}} B^{2}\label{eq:bhhx1}
\end{equation}
where
\begin{equation}
\beta_{K_{sp}^{pp}}=K_{sp,0}^{pp}\left(\frac{el_{p,0}^2}{4\hbar}\right)^2\label{eq: blh}
\end{equation}
is the diamagnetic shift coefficient of the same particle (hole-hole or electron-electron) exchange interaction. 
%Here $K_{sp,0}^{pp}$ is the value of this interaction energy in a zero magnetic field.

Using Eq.~\ref{eq:JE} and Eq.~\ref{eq:alpha}, one obtains an expression for the value of the particle-particle exchange diamagnetic shift coefficient, $\beta_{K_{sp}^{pp}}$: 
%that depends only on the zero magnetic field value of the hole-hole exchange interaction, $K_{sp,0}^{hh}$ and the relative dielectric constant of the QD material, $\epsilon_r$,
%expression for the field dependence of the hole-hole exchange energy
\begin{equation}
\beta_{K_{sp}^{pp}}=\frac{e^{10}}{2^{22}\pi^2 \hbar^2 \epsilon_0^4\epsilon_r^4\left(K_{sp,0}^{pp}\right)^3}. \label{eq:theortical_beta}
\end{equation}
Eq.~\ref{eq:theortical_beta} is important, since both the hole-hole exchange interaction  $K_{sp,0}^{hh}$ and its diamagnetic shift $\beta_{K_{sp}^{hh}}$
can be directly deduced from our measurements, providing thus a way to uniquely determine $\epsilon_r$. 

Since by Eq.~\ref{eq:eff_length_Bapp} the field dependence of the effective hole and electron lengths are different, $\gamma$ and $\eta$ are also field dependent: 

 \begin{equation}
\gamma(B)=\frac{l_e(B)}{l_h(B)}\approx\gamma_0 \left[ 1+\left(1-\gamma_0^4 \right)\left(\frac{e l_{h,0}^2}{4\hbar}\right)^2 B^2 \right], \label{eq:gammaB}
\end{equation}
where $\gamma_0$ and $ l_{h,0} $ are defined in the absence of magnetic field and we keep only lowest order terms in $B$. 
Using $\eta_0=\gamma_0-1$,  keeping only first order terms in $\eta_0$ one gets: 
 \begin{equation}
\eta(B)= \eta_0\left[1-4\left(\frac{el_{h,0}^2}{4\hbar}\right)^2B^2\right].\label{eq:etaB}
\end{equation}
Therefore, the field dependence of the product $\eta\cdot K^{hh}_{sp}$ up to lowest order terms in $B$, used in Eq.~\ref{eq:eta1} and Eq.~\ref{eq:E_BE} 
is given by 
 \begin{equation}
\eta(B)\cdot K^{hh}_{sp}(B)\approx\eta_0K^{hh}_{sp,0}-3\eta_0\beta_{K^{hh}_{sp}}B^2.
\end{equation}

To conclude this section, we note that from the measured hole-hole exchange term $K_{sp,0}^{hh}$ and its diamagnetic shift $\beta_{K^{hh}_{sp}}$, one can straightforwardly calculate, using Eq.~\ref{eq: blh}, the extent of the hole parabolic confining potential $l_{h,0}$. Then, by using Eq. \ref{eq:KEHR1}, the intrinsic short-range electron hole exchange term $E^{eh}_{sr}$, can be estimated as well.
Together with the measured value of $\eta_0$,  as we show below, the diamagnetic shifts of practically all the measured optical transitions, can be estimated.   

\subsection{The magnetic field dependence of the electron-hole exchange terms} 

The isotropic electron-hole exchange term, $K_{ss}^{eh}$, which equals half of the DE - BE fine structure splitting ($\delta_0^{1e1h}$, Table \ref{tab:g-factors}) is, as mentioned above, mostly due to the short-range exchange interaction~\citep{Takagahara2000, Ivchenko2005}. Its dependence on the magnetic field results as shown in Eq. \ref{eq:KEHR1} from its inverse dependence on the term $(l_h^2+l_e^2) =l_h^2(1+\gamma^2) \approx 2l_h^2(1+\eta)$ (or roughly twice the QD area~\citep{Takagahara2000}). 
Therefore, by using the field dependence of $l_h$ (Eq. \ref{eq:eff_length_Bapp}) and that of $\eta$ (Eq. \ref{eq:etaB}) in Eq.~\ref{eq:KEHR1}, while keeping only lowest terms in $B^2$ and $\eta_0$ one obtains: 
\begin{equation}
K_{ss(p)}^{eh}(B)=K_{ss(p),0}^{eh}+\beta_{K^{eh}_{ss(p)}}B^2; \label{eq:KEHB}
\end{equation}
where 
\begin{equation}
\beta_{K^{eh}_{ss}}= 2(1+2\eta_0)\frac{K_{ss,0}^{eh}}{K_{sp,0}^{hh}}\beta_{K^{hh}_{sp}}
,\label{eq:BKEHB1}
\end{equation}
and  
\begin{equation}
\beta_{K^{eh}_{sp}}= 2\frac{K_{sp,0}^{eh}}{K_{sp,0}^{hh}}\beta_{K^{hh}_{sp}}
,\label{eq:BKEHB2}
\end{equation}
where we also used Eq.~\ref{eq: blh}. 

Eqs. \ref{eq:BKEHB1} and \ref{eq:BKEHB2} relate the field dependence of the electron-hole exchange to the hole-hole exchange and thus provide a way for 
estimating the magnetic field dependencies of the electron hole exchange interactions. 
We use these relations to compare the model estimate with the actually measured diamagnetic shifts. 

\subsection{The diamagnetic shifts of the bright and dark excitons }

The diamagnetic shift of the free exciton is proportional to its wavefunction area \citep{Ivchenko2005}
in a plane normal to the direction of the magnetic field:

\begin{equation}
\beta_{X^{0}_{BE}}=\frac{e^{2}}{8\mu}\bravket f{\hat{\rho_{ex}}^{2}}f\label{eq: Exciton area}
\end{equation}

%where $\mu$ is the reduced exciton mass, 
%$\hat{\rho}$ is the electron-hole relative coordinate 
%and $\bravket{\hat{\rho}^{2}}$ is the expectation value of  $\hat{\rho}^{2}$.
Here, $\rho_{ex}=\rho_e-\rho_h$ is the relative radius vector between the electron
and hole, $f\left(\rho_{ex}\right)$ is the excitonic envelope wavefunction,
and $\mu=m_{e}m_{h}/\left(m_{e}+m_{h}\right)$ is the 
reduced mass of the electron and hole. 

For a confined exciton the effect of the geometric confinement potential on the exciton center of mass 
is approximated by anisotropic deformation of the exciton wave function $f(\rho_{ex})$, which is  taken to be a 
hydrogeniclike ellipsoid of revolution, characterized by effective
anisotropic Bohr radii parallel to its three principal axes. Equivalently, since the Bohr radii are inversely proportional to the reduced mass, one uses in this approximation three direction-dependent reduced exciton masses ($\mu_i, i=x,y,z$). For such an asymmetric hydrogenic wave function, the diamagnetic shift
in Faraday configuration, with magnetic field along $z$, is given by first order perturbation theory \citep{Taguchi1988,Bayer1998} as:
\begin{equation}
\beta_{X^{0}_{BE}}=\frac{4\pi^2\hbar^4\epsilon_0^2\epsilon_r^2}{e^{2}\mu\mu_x\mu_y} ; \frac{1}{\mu}=\frac{1}{3}\left(\frac{1}{\mu_x}+\frac{1}{\mu_y}+\frac{1}{\mu_z}\right)\label{eq: Dia ExcShift}
\end{equation}
While the excitonic binding energy at zero magnetic field is given by the regular Rydberg formula:
\begin{equation}
E_{BE}=\frac{e^4\mu}{32\pi^2\hbar^2\epsilon_0^2\epsilon_r^2}\label{eq: Exc Binding}
\end{equation}

After the exciton recombination, the QD remains empty of carriers. This final ``vaccum'' state does not shift with the magnetic field and therefore the measured diamagnetic shift of the BE transition is positive and given by Eq.~\ref{eq: Dia ExcShift}. 

In our circularly symmetric QD model, $\mu_x=\mu_y=\mu_{xy}$ and $\mu_z=\frac{\mu_{xy}}{a}$, 
where $a=l_z/l_{xy}$ is the aspect ratio between the normal and lateral dimensions of the QD confined excitonic wavefunction. 
One can therefore express  $\mu=\chi\mu_{xy}$, where $\chi= \frac{3}{2+a}$. In the 2D limit $a=0$, $\mu_z=\infty$ and $\chi=3/2$. 

Using Eq.~\ref{eq: Dia ExcShift} and Eq.~\ref{eq: Exc Binding} one can now express 
the excitonic diamagnetic shift in terms of the excitonic binding energy:
\begin{equation}
\beta_{X^{0}_{BE}}=\frac{\chi^2 e^{10}}{2^{13}\pi^4\hbar^2\epsilon_0^4\epsilon_r^4}\frac{1}{E_{BE}^3}\label{eq:DiaBinding}
\end{equation}

From Eq.~\ref{eq:E_X0DE} and the quadratic field dependence of all the direct Coulomb and exchange terms it immediately follows that: 
\begin{equation}
\beta_{X^{0}_{DE}}=\beta_{X^{0}_{BE}}-2\beta_{K^{eh}_{ss}}\label{eq:DiaDark}
\end{equation}
In a similar way one can calculate the expected diamagnetic shifts of the charged excitons as well (see Table \ref{tab:beta summary}, below).

To conclude this section we note that   
Eq.~\ref{eq:DiaBinding} much resembles Eq.~\ref{eq:theortical_beta}. In fact if one uses $J^{eh}_{ss}=(4-2\eta_0)K^{hh}_{sp,0}$ (Eq.~\ref{eq:E_BE}) as an estimate for the excitonic binding energy ($E_{BE}$), the ratio between the diamagnetic shifts of the hole-hole exchange interaction and that of the bright exciton depends only on the geometrical ratios $\chi$ and $\eta_0$: 
\begin{equation}
\left(\frac{\beta_{X^0_{BE}}}{\beta_{K^{hh}_{sp}}}\right)\approx\frac{8\chi^2}{\pi^2(1-6\eta_0)} ; \lim_{\chi,\eta_0 \longrightarrow \frac{3}{2}, 0}\left(\frac{\beta_{X^0}}{\beta_{K^{hh}_{sp}}}\right)=2\left(\frac{3}{\pi}\right)^2\label{eq:Xilim}
\end{equation}
Surprisingly, for equally confined electron and hole in the 2D limit, the model predicts a universal ratio which is independent of the QD dimensions and material properties.    

\section{Discussion}

\subsection{Estimating the hole-hole and electron-hole exchange terms}

The exchange terms can now be estimated directly from the measured PL spectrum of the quantum dot.

From Eq.~\ref{eq:S0} it directly follows that 
\begin{eqnarray}
K_{sp,0}^{hh}=(E_{X_{T_{0}}^{+2}}-E_{X_{S_{0}}^{+2}})/2=2.79(15)~\text{meV},\label{eq:S0M} 
\end{eqnarray}
(see Figures \ref{fig:MPL}(b) and \ref{fig:Xp2-MPL}).
%The hole-hole exchange $K_{sp,0}^{hh}$ is given by half the energy difference between the $XX^{+2}_{T_0}$ and the $XX^{+2}_{S_0}$ spectral lines.
From Eq.~\ref{eq:E_X0DE} it follows that
\begin{equation}
K_{ss,0}^{eh}=(E_{X^{0}_{BE}}-E_{X^{0}_{DE}})/2=0.135(5)~\text{meV},\label{eq:E_X0DEM}
\end{equation}
(see Figure \ref{fig:BE-DE-MPL}).
From Eq.~\ref{eq:KEHR} one expects that $K_{sp,0}^{eh}=0.5K_{ss,0}^{eh}\approx63(5)\mu eV$.
This result should be compared with Eq.~\ref{eq:T3} from which it follows that 
%\begin{equation}
$K_{sp,0}^{eh}= (E_{X^{+2}_{T_0}}- E_{X^{+2}_{T_3}})/2\approx0(20)~\mu\text{eV}$, (see Figures \ref{fig:MPL}(b) and \ref{fig:Xp2-MPL}). 
Unfortunately, there is a small discrepancy here, probably because of the lack of spectral resolution, due to the spectral lines' widths and a possible degeneracy removal between the $T_0$ and $T_{\pm 3}$ two-heavy-hole triplet states \citep{Ediger2007}, which our model does not take into account.  
%=0.063(5) meV,\label{eq:T3M}
%\end{equation}

From Eq.~\ref{eq:E_X+} it follows that
 \begin{eqnarray}
J_{ss}^{hh}-J_{ss}^{eh}=E_{X^{+1}}-E_{X^{0}_{DE}}=-130(15)~\mu\text{eV}.\label{eq:E_X+M}
 \end{eqnarray} 
%and that from Eq.\ref{eq:T0} it follows that
%\begin{eqnarray}
%J_{sp}^{hh}-J_{sp}^{eh}=E_{X^{+1}}-E_{X_{T_{0}}^{+2}}=-0.03(2) meV,\label{eq:T0M} 
%\end{eqnarray}
(see Figures \ref{fig:BE-DE-MPL} and \ref{fig:MPL}(a) ).
Likewise, from Eqs.~\ref{eq:T0} 
%and \ref{eq:eta1} it follows that:
\begin{equation}
J_{sp}^{hh}-J_{sp}^{eh}=E_{X_{T_{0}}^{+2}}-E_{X^{+1}}=-30(15)~\mu\text{eV},\label{eq:T0M}
%=J_{sp}^{hh}-J_{sp}^{eh}=(J_{ss}^{hh}-J_{ss}^{eh})/4 
\end{equation}
(see Figures \ref{fig:MPL}(a) and \ref{fig:Xp2-MPL})
which within the limited resolution of our measurements agrees well with the expected ratio of 1:4 between $J_{sp}^{hh}-J_{sp}^{eh}$ and $J_{ss}^{hh}-J_{ss}^{eh}$, following Eq.~\ref{eq:eta1}. 
 
We  proceed by using Eqs.~\ref{eq:JE} and \ref{eq:eta1} to 
get an estimate for $\eta_0$:
\begin{equation}
\eta_0=(J_{ss}^{hh}-J_{ss}^{eh})/2K_{sp,0}^{hh}=-0.024(3),\label{eq:etaM} 
\end{equation}
Likewise, using Eq.~\ref{eq:E_BE} we get an estimate for the binding energy of the BE: 
\begin{equation}
E_{BE}\approx J_{ss}^{eh}\approx 4K_{sp,0}^{hh}(1-\frac{1}{2}\eta_0)=11.3(4)~\text{meV}.\label{eq:E_BEM}
\end{equation}

\subsection{The measured and estimated diamagnetic shifts}

The measured diamagnetic shifts are summarized in the second column of Table \ref{tab:beta summary}. 
The diamagnetic shifts of the spectral lines are displayed in the upper 6 rows of the table. 
From these measurements, we directly obtain the measured diamagnetic shifts of the hole-hole and electron-hole exchange terms:  
Using  Eq.~\ref{eq:S0} we get: 
\begin{equation}
\beta_{K_{sp}^{hh}}=(\beta_{X_{T_{0}}^{+2}}-\beta_{X_{S_{0}}^{+2}})/2
\end{equation}
using Eq.~\ref{eq:DiaDark} we get:
\begin{equation}
\beta_{K^{eh}_{ss}}=(\beta_{X^{0}_{BE}}-\beta_{X^{0}_{BE}})/2,
\end{equation}
and using Eq.~\ref{eq:T3} we get:
\begin{equation}
\beta_{K_{sp}^{eh}}=(\beta_{E_{X_{T_{0}}^{+2}}}-\beta_{E_{X_{T_{\pm3}}^{+2}}})/2. 
\end{equation}
The obtained values are also listed in the last 3 rows of Table \ref{tab:beta summary}. 

\begin{table}
\begin{centering}
\begin{tabular}{cccc}
\toprule 
 & Measured & Calculated & Theoretical expression\tabularnewline
\midrule
\midrule 
\textbf{$ \beta_{X^0_{BE}}  $} & $8.44$ &$8.44$  & Eq.~\ref{eq:Xilim} ($\chi=1.36$) \tabularnewline
%\midrule 
\textbf{$ \beta_{X^0_{DE}}  $} & $7.0$ &  $7.2$ & $ \beta_{X^0_{BE}} -2\beta_{K^{eh}_{ss}}$  \tabularnewline
%\midrule 
\textbf{$ \beta_{X^{+}}  $} & $7.85$ & $7.95$ & $\beta_{X^0_{DE}}-6\eta_{0}\beta_{K^{hh}_{sp}}  $\tabularnewline % From Ayal Thesis table 4.2
%\midrule 
\textbf{$ \beta_{X^{+2}_{T_{0}}}  $ } & $7.6$ & $8.09$ & $\beta_{X^{+}}-\frac{3}{2}\eta_{0}\beta_{K^{hh}_{sp}}$ \tabularnewline
%\midrule
\textbf{$ \beta_{X^{+2}_{S_{0}}}  $ } & $-5.8$ & $-5.6$ & $\beta_{X^{+2}_{T_{0}}}-2\beta_{K^{hh}_{sp}}$ \tabularnewline
%\midrule  
\multirow{1}{*}{\textbf{$ \beta_{X^{+2}_{T_{\pm 3}}}  $ }} & \multirow{1}{*}{$6.9$} & \multirow{1}{*}{$6.9$} & $\beta_{X^{+2}_{T_{0}}}-2\beta_{K^{eh}_{sp}}$ \tabularnewline
%\midrule  
\textbf{$ \beta_{K^{hh}_{sp}}  $} & $6.7$ &$6.7$  & Eq.~\ref{eq:theortical_beta} ($\epsilon_r=14.24$) \tabularnewline
%\midrule  
\textbf{$ \beta_{K^{eh}_{ss}}  $} & $0.62$ &$0.62$  &Eq.~\ref{eq:BKEHB1} \tabularnewline
%\midrule 
\textbf{$ \beta_{K^{eh}_{sp}}  $} & $0.55$ &$0.35$  & Eq.~\ref{eq:BKEHB2}  \tabularnewline
\midrule 
\bottomrule
\end{tabular}
\par\end{centering}
\caption{Measured and modeled diamagnetic shifts ($\beta$ values) in $\mathrm{\mu eV/T^2}$. The estimated relative errors are about $5\%$. For the model we used the zero field measured $K^{hh}_{sp,0}=2.79(15)~\text{meV}$ (Eq.~\ref{eq:S0M}), $K^{eh}_{ss,0}=0.5\delta_0^{11}=0.135(7)~\text{meV}$ (Table \ref{tab:g-factors}) and $\eta_{0} =  -0.024(3)$ (Eq.~\ref{eq:etaM}).  \label{tab:beta summary} }
\end{table}

The calculated diamagnetic shifts are displayed for comparison in the third column of Table \ref{tab:beta summary}. 
The expressions used for these calculations were developed in Section~\ref{sec:Theoretical Analysis} using the Hartree-Fock approximation and the cylindrical parabolic potential model for the QD, are displayed and referenced in the fourth column of the table. 
As one can see, the agreement is surprisingly decent, despite the simplicity of our model for the QD potential and despite the fact that the Hartree-Fock model completely ignores the correlation terms in the Coulombic interactions \cite{Regelman-Zunger} and we do not include any single-particle reshaping that occurs from interaction. 

\subsection{The QD dielectric constant and average composition }

By substituting the measured $K_{sp,0}^{hh}$ and $\beta_{K_{sp}^{hh}}$ in Eq.~\ref{eq:theortical_beta} we find that the average dielectric constant of the QD is $\epsilon_r=14.24(4)$. 
%We show below, how this value can be used to determine the average composition of the ternary alloy of the QD.
This experimentally deduced value provides a way to estimate the QD effective composition. 
Since the QD comprises of two binary semiconductors, GaAs and InAs, one may interpolate the value of any material property (Q) of the ternary material 
using an average effective composition $x$ for the QD ternary material In$_{x}$Ga$_{1-x}$As  using quadratic interpolation formula: 
\begin{equation}
 Q(x)=xQ^{In}+(1-x)Q^{Ga}-c^Qx(1-x)\label{eq:quad}
 \end{equation} 
here, $Q^{In}$ ($Q^{Ga}$) is the Q value of the binary material InAs (GaAs) and $C^Q$ is a bowing parameter characterizing Q for the ternary material. 
The material parameters that we use for the dielectric constants are given in Table \ref{tab:Roth}, 
in which we list and reference the relevant input parameters for the quadratic interpolations used in this work.
 
The effective QD composition which results in the experimentally deduced dielectric constant is $x=0.70\pm0.05$.
Interestingly, similar value for $x$ is obtained if one interpolates the measured electronic QD $g$-factor. 

The isotropic electronic $g$-factor in bulk semiconductors can be
analytically calculated by the Roth's formula \citep{Roth1959}:
\begin{equation}
g_{e}=2-\frac{2}{3}\frac{E_{p}\Delta}{E_{g}(E_{g}+\Delta)}\label{eq:roth}
\end{equation}
where $E_{g}$ is the band gap energy between the valence and conduction
bands, $\Delta$ is the split-off gap (between the valence band and
the spin-orbit band) at $k=0$, and $E_{p}$ is the Kane energy defined
as $\mbox{\ensuremath{E_{p}\equiv\frac{2\hbar^{2}}{m}}|\ensuremath{\bra s\pa x\ket x}\ensuremath{|^{2}}}$
, where $\ket s$ and $\ket x$ are the crystal Bloch functions of
the electron in the conduction band and in one of the three p-like
degenerate valence bands, respectively. 
The quantum confinement due to the QD potential breaks the periodicity of the electronic wavefunctions,
and the derivation of the Roth's formula collapses \citep{Pryor2006}. Nevertheless, as long
as the confinement energy is much smaller than the energies $\Delta$,
$E_{p}$ and $E_{g}$, one still expects Roth's formula to be a qualitative approximation that provides an upper bound on the band gap (a lower bound of the composition $x$).
Indeed, the typical separation between the confined carriers' energy levels in
our QD is of order $10-30~\text{meV}$ \citep{Benny2011a}, much smaller
than $\Delta$, $E_{p}$ and $E_{g}$ (see Table \ref{tab:Roth}).

Therefore, we proceed by interpolating the values of $\Delta$ and $E_{p}$ 
using Eq.~\ref{eq:quad} and the material parameters in Table \ref{tab:Roth}. 
For the QD band gap, $E_{g}^{QD}$, we use the directly measured value of the $X_{BE}^{0}$ spectral line, as it takes into account the
quantum confinement and lattice mismatch strain effects \citep{Bree2012}. 
In order to get better estimate for the QD  bandgap, however, one has to add to the measured PL the excitonic binding energy. As discussed above, we use the direct Coulomb term $J^{eh}_{ss}$ (Eq.~\ref{eq:E_BE}), as an estimate for the excitonic binding energy. Thus $E_g^{QD}=E^{X^0_{BE}}+J^{eh}_{ss}$, where $J^{eh}_{ss}\approx 11.3~\text{meV}$.
The Roth' formula for the QD is therefore given by:
\begin{equation}
g_{e}(x)=2-\frac{2}{3}\frac{E_{p}(x)\Delta(x)}{E_{g}^{QD}(E_{g}^{QD}+\Delta(x))}\label{eq: g(x)}
\end{equation}

Using the measured value $g_{e}(x)=-0.55$ in Eq.~\ref{eq: g(x)} and the material parameters from Table \ref{tab:Roth} we get an estimate for the effective QD composition $x=0.76\pm 0.04$, where the uncertainty in $x$ includes also the uncertainties in the tabulated band parameters.
This value agrees to within the experimental uncertainties with the value obtained from the dielectric constant deduced from the diamagnetic shifts.

\begin{table}
\begin{centering}
\begin{tabular}{cccccc}
\toprule 
 & GaAs & InAs & Bowing &    best x &   In$_{x}$Ga$_{1-x}$As\tabularnewline
\midrule
\midrule 
$\epsilon_{r}$ & $15.15^a$ & $12.46^a $& $0.67^b$ & 0.71 & 14.24\tabularnewline

%\midrule 
$E_{p}[eV]$ & $28.8^c$ & $21.1^c$ & $-1.48^c$ & 0.76 & 23.22\tabularnewline
%\midrule 
\textbf{$E_{g}[eV]$} & $1.519^c$ & $0.418^c$ & $2.22^c$ & 0.76 &  $1.295^{*}$\tabularnewline
%\midrule 
\textbf{$\Delta[eV]$} & $0.341^c$ & $0.39^c$ & $0.15^c$ & 0.76 & 0.351\tabularnewline
%\midrule 
\textbf{$g_{e}$ }Calculated & -0.317 & -14.3 &  & 0.76 & $-0.55$\tabularnewline
%\midrule 
\multirow{1}{*}{\textbf{$g_{e}$ }Measured} & $\multirow{1}{*}{-0.484}^d$ & $\multirow{1}{*}{-14.9}^d$ &  & & $-0.55$\tabularnewline
\midrule 
\bottomrule
\end{tabular}
\par\end{centering}
\caption[The average  In$_{x}$Ga$_{1-x}$As ternary QD composition.]{The estimated average composition of the QD estimated from the deduced dielectric constant and the measured electronic $g$-factor. The dielectric constant and band parameters are 
interpolated using Eq.~\ref{eq:quad} and the tabulated parameters. * the measured excitonic emission+excitonic binding energy (see text).
a Ref.\citep{Adachi1999}; b Ref.\citep{Goldberg1996}; c Ref. \citep{Ram-Mohan2001}; d Ref.\citep{Ivchenko2005}.    \label{tab:Roth}}
\end{table}

\subsection{The QD dimensions}

One can use the magneto-PL spectroscopy to estimate the QD dimensions.
By substituting the measured $\beta_{K_{sp}^{hh}}$ in  Eq.~\ref{eq: blh} one can find the extent of the hole wavefunction as given by Eq.~\ref{eq:WF1}. 
Using the experimentally estimated $\gamma_0=0.976(2)$  the electron wavefunction extent can be obtained as well.
This way we find that $l_h$ and $l_e$ are $11.7(3)$ and $11.4(3)~\text{nm}$, respectively.  

By substituting the measured ratio between the diamagnetic shifts of the hole-hole exchange interaction $\beta_{K_{sp}^{hh}}$ and that of the BE $\beta_{X^0_{BE}}$ in Eq.~\ref{eq:Xilim} we obtain $\chi=1.36(2)$, which is slightly less than 1.5 expected for a truly 2D exciton. 
To account for this discrepancy we consider the extent of the excitonic wavefunction in the $z$ direction $l_z$. This may be attributed to the fact that the electron mass is isotropic, unlike that of the HH which is much heavier in the $z$ direction \cite{Ivchenko2005}. The aspect ratio $a$ between the extent of the BE wavefunction in the $z$ direction to its extent in the plane is given by $3/\xi-2=0.20(1)$ which leads to $l_z\approx 2.3(1)~\text{nm}$. 

It follows that the estimated dimensions of the QD are about 23~nm in diameter and 4.5~nm in height. 
We note that the lateral dimensions are probably underestimated. 
This is because the confining length of the parabolic potential well model ($l_p$) is about a factor of 2 smaller than that of an infinite 2D potential well model which produces the same energy difference between the s- and p-shells of the confined carrier ($\hbar\omega_p$ as defined in Eq.~\ref{eq: Harmonic_length1}). The estimated QD height is probably slightly overestimated, due to the penetration of the electronic wavefunction  into the GaAs binary barriers. We note here that the extent of the electronic wavefunction along the growth direction leads to the following relations between the direct Coulomb terms $\abs{J^{ee}_{ss}}< \abs{J^{hh}_{ss}}\approx\abs{J^{eh}_{ss}}$. These relations leads in turn to the quite general experimental observation that the negatively charged exciton spectral line ($X^{-1}$)  is a few meV lower in energy than the positively charged  exciton ($X^{+1}$) line, while the later is quite close in energy to the neutral exciton line ($X^{0}_{BE}$, see Fig. \ref{fig:MPL}a).          

\section{Summary}
We experimentally investigated using polarization sensitive magneto-PL spectroscopy a well-characterized  In$_{x}$Ga$_{1-x}$As QD in the Faraday configuration. We systematically measured the Zeeman splittings of neutral, singly and doubly charged excitons. The $g$-factors of the bright and dark excitons were measured first and their values were used to show that  the Zeeman splittings of various charged excitonic lines can be quite well described by a simple arithmetic model resulting from sums and differences of the $g$-factors of the confined electron and holes in their respective energy levels. In particular, from these measurements we extracted the $g$-factor of the hole in its second confined energy level and showed that it has opposite sign with respect to the hole in its first energy level. 

The measured diamagnetic shifts of the excitonic transitions were carefully measured as well. All the transitions showed quadratic dependence on the magnitude of the externally applied magnetic field. In particular, we observed a pronounced negative diamagnetic shift of one of the sptectral lines for transitions from the doubly positively-charged excitons ($X_{S_0}^{+2}$). The magneto-PL measurements were all quantitatively explained using a Hartree-Fock model to describe the direct Coloumb and exchange interactions between up to 4 confined carriers in the QD.

We used a two dimensional cylindrically symmetric parabolic potential model in order to analytically calculate the Coulomb and exchange integrals and their magnetic field dependence.
The model quantitatively describes the measured diamagnetic shifts of many excitonic transitions and it accurately describes the hole-hole and electron-hole direct Coulomb and exchange interactions and their magnetic field dependence.  
From our measurements and model we obtained an estimate for the QD average dielectric constant ($\epsilon_r$), and for its lateral dimensions $l_h\approx l_e$. In addition, we show that while the 2D model is adequate for describing the confined heavy hole wavefunctions, the perpendicular extent of the electron wavefunction must be considered in order to quantitatively account for the exciton diamagnetic shift. The latter extent provides an estimate for the QD height, in decent agreement with the structural data at hand. Lastly, we show that by interpolating both the QD electronic $g$-factor and its dielectric constant between the QD's binary constituents GaAs and InAs we succeed to provide similar estimates for the  In$_{x}$Ga$_{1-x}$As QD average composition $x$. Although simple, the model provides a good understanding of the experimental magneto optics of charge-tunable quantum dots.
\medskip{}

\begin{acknowledgments}
We thank Dr. J. Tilchin, and Professor E. Lifshitz for their help, and Professors M. M. Glazov, and  E. L.
Ivchenko for valuable discussions. 
The support of the Israeli Science Foundation (ISF), and that of the German Israeli Research
Cooperation\textemdash DIP (DFG-FI947-6-1) are gratefully acknowledged. 
\end{acknowledgments}

\bibliographystyle{apsrev4-2}
\phantomsection\addcontentsline{toc}{section}{\refname}\bibliography{Bibilography}

\begin{thebibliography}{60}
\bibitem [{NIS()}]{NISTdisclaimer}%
  \BibitemOpen
  \href@noop {} {}\bibinfo {note} {Certain commercial equipment, instruments,
  software or materials are identified in this paper to specify the procedures
  or equipment used. Such identification is not intended to imply
  recommendation or endorsement by the National Institute of Standards and
  Technology, nor is it intended to imply that the materials, equipment or
  software identified are necessarily the best available for the
  purpose.}\BibitemShut {Stop}%
\end{thebibliography}

\end{document}